\documentclass[12pt]{article}
\usepackage{graphicx, amsmath, amssymb}
\usepackage{hyperref}
\usepackage{epstopdf}

\newcommand{\half}{{\frac{1}{2}}}
\newcommand{\mbf}[1]{\mathbf{#1}}

\begin{document}

\begin{flushright}
{\small
SLAC--PUB--14381\\
\date{today}}
\end{flushright}

\vspace{30pt}

\centerline{\LARGE  {Applications of AdS/QCD and Light-Front Holography}}

\vspace{5pt}

\centerline{\LARGE {to Baryon Physics }}

\vspace{20pt}

\centerline{{
Stanley J. Brodsky,$^{a}$ and
Guy F. de T\'eramond$^{b}$
}}

\vspace{20pt}

{\centerline {$^{a}${SLAC National Accelerator Laboratory, 
Stanford University, Stanford, CA 94309, USA}}

\vspace{4pt}

{\centerline {$^{b}${Universidad de Costa Rica, San Jos\'e, Costa Rica}}

 \vspace{40pt}

\begin{abstract}

The correspondence between theories in anti--de Sitter space  and  field theories in physical space-time leads to an analytic, semiclassical model for strongly-coupled QCD which has scale invariance at short distances and color confinement at large distances.   These equations, for both mesons and baryons, give a very good representation of the observed hadronic spectrum, including a zero mass pion.  Light-front holography allows hadronic amplitudes in the AdS fifth dimension to be mapped to frame-independent light-front wavefunctions of hadrons in physical space-time, thus providing a  relativistic description of hadrons at the amplitude level.  The meson and baryon wavefunctions derived from light-front holography and AdS/QCD also have remarkable phenomenological features, including predictions for the  electromagnetic form factors and decay constants.  The approach can be systematically improved using light-front Hamiltonian methods.  Some novel features  of QCD for baryon physics are also discussed.

\end{abstract}

\newpage

\section{Introduction}

The light-front wavefunctions (LFWFs) of relativistic bound states in quantum chromodynamics  provide a fundamental description of the structure and internal dynamics of hadronic states in terms of their constituent quark and gluons  at a fixed light-front time  $\tau = t + z/c$, the time marked by the
front of a light wave,~\cite{Dirac:1949cp,Brodsky:1997de} --  rather than at instant time $t$, the ordinary time. Unlike instant time quantization, the Hamiltonian equation of motion in the light front (LF) is frame independent. The simple structure of the light-front vacuum allows an unambiguous
definition of the partonic content of a hadron in QCD and of hadronic light-front wavefunctions
which relate its quark
and gluon degrees of freedom to their asymptotic hadronic state.  For example, the proton's eigenstate $\vert p\rangle$, the lightest $Q=+1$, $B=+1 $ eigenstate of the QCD LF Hamiltonian,  can be expanded in terms of the light front  Fock components: $\psi_n(x_i, \mbf{k}_{\perp i}, \lambda_i)$ corresponding to its $\langle p \vert uud \rangle,  
\langle p \vert uudg\rangle , \langle p \vert uudQ\bar Q\rangle$, etc.  projections.
Here $x_i = k^+_i / P^+ =(k^0_i+ k^z_i)/(P^0+P^z)$ are the light-front momentum fractions of the constituents. The plus momentum and transverse momenta are  conserved at fixed $\tau$, $\sum^n_{i=1} x_i =  1$  and $\sum^n_{i=1} \mbf{k}_{\perp i} =  1$ in each $n-$parton wavefunction.  Remarkably, the 
$\psi_n(x_i, \mbf{k}_{\perp i}, \lambda_i)$ are independent of the hadrons 4-momentum $P^\mu.$  In light-cone gauge $A^+=0$, the gluon quanta have $S^z= \pm 1$ and there are no ghosts. Angular momentum on the LF is simply 
$J^z= \sum^{n}_{i=1} S^z_i + \sum^{n-1}_{i=1} L^z_i$ since there are only $n-1$ independent orbital angular momentum.  The angular momentum carried by the gluon  in the proton as measured by experiment is  simply the mean value of $S^z_g+ L^z_g$ summed over Fock states.  The structure functions measured by experiment  in deep inelastic proton scattering (DIS) are related to the absolute square of the LFWFs summed over Fock states. The proton's Dirac and Pauli form factors have an exact representation as the spin-conserving and spin-flip overlaps of initial and final wavefunctions, respectively. Unlike the ordinary instant form, there are no diagrams where the current couples to vacuum currents, and the boost of the LFWFs is trivial.   Given, the LFWFs, one can compute hadronization at the amplitude level from the coalescence of a set of comoving color-singlet quarks and gluons,~\cite{Brodsky:2008tk} in analogy to the formation of moving atoms in quantum electrodynamics.~\cite{Munger:1993kq}

There are many novel features of baryons which are illuminated using the LFWF representation:

 \begin{enumerate}

 \item The higher Fock states of baryons contain intrinsic heavy quark~\cite{Brodsky:1980pb}  such as $\vert uud c\bar c \rangle$; these can arise from gluon splitting, but also from multi-connected amplitudes. The intrinsic heavy quark  (IQ) contributions are maximal at minimum off-shellness of the LFWF; i.e, when the partons have the same rapidity. This corresponds to 
$x_i  \propto \sqrt{\mbf{k}^2_{\perp i}+m^2_i}$, so that the heavy quark have most of the momentum. In a collision the states are materialized and the quarks with the same rapidity such as $\vert Q ud \rangle$ coalesce to high $x_F$ baryons such as the   $\Lambda_c(cud)$ 
and the $\Lambda_b(bud)$ observed at high $x_F$ at the ISR.~\cite{Bari:1991ty}  SELEX~\cite{Mattson:2002vu} has also discovered  double charm baryons 
$\vert ccu \rangle$ and $\vert ccd \rangle$ at high $x_F$ this way.   This suggests using the LHC beam in a fixed target mode to observe very heavy baryons at high $x_F$ such as the $\Omega_b(bbb)$.~\cite{Lansberg}  Intrinsic heavy quarks also provide a novel mechanism to produce the Higgs meson and $Z^0$ at high $x_F$.~\cite{Brodsky:2006wb,Brodsky:2007yz}

\item High $p_T$ hadrons can be created directly from hard subprocesses such as $g q \to M q$ and $q \bar q \to B + \bar q$, rather than from quark of gluon fragmentation.~\cite{Arleo:2009ch}  Since the direct hadrons are produced with small transverse size, they are color transparent and can traverse the nuclear medium without absorption.   The direct processes have been seen in   many experiments. They also account for the remarkable baryon anomaly observed in ion-ion collisions,~\cite{Brodsky:2008qp} as well as the anomalously large power-law falloff of  the inclusive cross sections $ {Ed\sigma\over d^3p}(p p \to H X)$  for meson and baryon production at fixed $\theta_{cm}$ and $x_T = 2 p_t/\sqrt s$ .  Since there are no same-side hadrons, the direct processes are energy efficient, requiring the minimum incident parton momentum fractions where the parton distributions are maximal.

\item The Sivers effect in deep inelastic lepton-polarized proton scattering is most easily computed as an interference  of the proton's  $L^z=0$ and $L^z=\pm 1$ 
LFWFs.~\cite{Brodsky:2002cx}  The Sivers spin-correlation $\vec S_p \cdot \vec p_q \times \vec q$   is $T-odd$, which  reflects the different phases of the  $L^z=0$ and $L^z=\pm 1$ amplitudes due to final state interaction of the scattered quark with the proton's spectators.  The Sivers correlation for each quark is thus seen to be  proportional to that quark's contribution to the proton orbital angular momentum. The Sivers correlation has the opposite sign in Drell-Yan reactions~\cite{Collins:2002kn,Brodsky:2002rv}  because it measures initial state scattering of the annihilating quark. There are many other novel factorization-breaking effects which arise due to initial state or final state scattering of the active quark with the spectators, such as the double Bohr-Mulders correlation which leads to a breakdown of the PQCD Lam-Tung relation in Drell-Yan reactions.~\cite{Boer:2002ju}

\item  The quark condensate, normally identified as a vacuum expectation value $\langle 0 \vert \bar \psi \psi \vert 0 \rangle$ is actually an ``in-hadron'' condensate 
$\langle 0 \vert \bar \psi \psi \vert H \rangle$ associated with the $q \bar q$ sea quark excitations in the hadron's higher particle number Fock states.~\cite{Brodsky:2009zd,Brodsky:2010xf} The QCD vacuum is trivial - equal to the vacuum of the free theory in the front form, and there is thus no contribution to the cosmological constant within this framework.
\end{enumerate}

\section{AdS/QCD and Light Front Holography}

The AdS/CFT correspondence~\cite{Maldacena:1997re} between
string states on anti--de Sitter (AdS) space-time and conformal gauge field theories (CFT) in physical space-time has
brought a  new set of tools for studying the dynamics of strongly coupled quantum field theories, and it has led
to new analytical insights into the confining dynamics of QCD which is difficult to realize using other methods. 
Most important, it provides 
an initial approximation to QCD which is 
analytically tractable and which can be systematically improved.   The original conformal theory can be modified in the far infrared region of AdS space, for example by the introduction of a dilaton background,  which yields a confining potential between the colored quarks. The resulting model is usually called AdS/QCD.

One of the most remarkable features of AdS/QCD is the connection between the description of hadronic modes in AdS space and
the Hamiltonian formulation of QCD in physical space-time quantized
on the light-front; i.e.,  at equal light-front time  $\tau$. The first step for establishing the correspondence of light-front QCD in physical 3+1 space with AdS space is to observe that the LF bound state Hamiltonian equation of motion in QCD has an essential dependence  in the invariant transverse variable $\zeta$,~\cite{deTeramond:2008ht} which measures the
separation of the quark and gluonic constituents within the hadron
at the same LF time.  The  variable $\zeta$ plays the role of the radial coordinate $r$ in atomic systems. The result is a single-variable light-front relativistic
Schr\"odinger equation. This first approximation to relativistic QCD bound-state systems is 
equivalent to the equations of motion that describe the propagation of spin-$J$ modes in a fixed  gravitational background asymptotic to AdS space.~\cite{deTeramond:2008ht}  The eigenvalues of the LF Schr\"odinger equation give the hadronic spectrum and its eigenmodes represent the probability amplitudes of the hadronic constituents.  By using the correspondence between $\zeta$ in the LF theory and $z$ in AdS space, one can identify the terms in the dual gravity AdS equations that correspond to the kinetic energy terms of  the partons inside a hadron and the interaction terms that build confinement.~\cite{deTeramond:2008ht}  The identification of orbital angular momentum of the constituents in the light-front  is also a key element in our description of the internal structure of hadrons using holographic principles.
This mapping was originally obtained
by matching the expression for electromagnetic current matrix
elements in AdS space with the corresponding expression for the
current matrix element using LF theory in physical space
time.~\cite{Brodsky:2006uqa}  More recently we have shown that one
obtains the identical holographic mapping using the matrix elements
of the energy-momentum tensor,~\cite{Brodsky:2008pf} thus providing
a consistency test and verification of holographic
mapping from AdS to physical observables defined on the light front.

\section{A SEMICLASSICAL LIGHT-FRONT APPROXIMATION TO QCD}
The eigenmass $M^2$ of hadrons in light-front theory is determined from the eigenvalue equation
\begin{equation}
\langle \psi(P') \vert P_\mu P^\mu \vert\psi(P) \rangle  = 
M^2  \langle \psi(P' ) \vert\psi(P) \rangle,
\end{equation}
where one can expand the initial and final hadronic state in terms of its Fock components. The computation is  simplified in the 
frame $P = \big(P^+, M^2/P^+, \vec{0}_\perp \big)$ where $P^2 =  P^+ P^-$.
We find
 \begin{equation} \label{eq:Mk}
 M^2  =  \sum_n  \! \int \! \big[d x_i\big]  \! \left[d^2 \mbf{k}_{\perp i}\right]   
 \sum_q \Big(\frac{ \mbf{k}_{\perp q}^2 + m_q^2}{x_q} \Big)  
 \left\vert \psi_n (x_i, \mbf{k}_{\perp i}) \right \vert^2  + {\rm (interactions)} ,
 \end{equation}
plus similar terms for antiquarks and gluons ($m_g = 0)$. The integrals in (\ref{eq:Mk}) are over
the internal coordinates of the $n$ constituents for each Fock state
\begin{equation}
\int \big[d x_i\big] \equiv
\prod_{i=1}^n \int dx_i \,\delta \Bigl(1 - \sum_{j=1}^n x_j\Bigr) , ~~~
\int \left[d^2 \mbf{k}_{\perp i}\right] \equiv \prod_{i=1}^n \int
\frac{d^2 \mbf{k}_{\perp i}}{2 (2\pi)^3} \, 16 \pi^3 \,
\delta^{(2)} \negthinspace\Bigl(\sum_{j=1}^n\mbf{k}_{\perp j}\Bigr),
\end{equation}
with phase space normalization
$\sum_n  \int \big[d x_i\big] \left[d^2 \mbf{k}_{\perp i}\right]
\,\left\vert \psi_n(x_i, \mbf{k}_{\perp i}) \right\vert^2 = 1$.

The LFWF $\psi_n(x_i, \mathbf{k}_{\perp i})$ can be expanded in terms of  $n-1$ independent
position coordinates $\mathbf{b}_{\perp j}$,  $j = 1,2,\dots,n-1$, 
conjugate to the relative coordinates $\mbf{k}_{\perp i}$, with $\sum_{i = 1}^n \mbf{b}_{\perp i} = 0$.  
We can also express (\ref{eq:Mk})
in terms of the internal impact coordinates $\mbf{b}_{\perp j}$ with the result
\begin{equation}   
 M^2  =  \sum_n  \prod_{j=1}^{n-1} \int d x_j \, d^2 \mbf{b}_{\perp j} \,
\psi_n^*(x_j, \mbf{b}_{\perp j})  \\
 \sum_q   \left(\frac{ \mbf{- \nabla}_{ \mbf{b}_{\perp q}}^2  \! + m_q^2 }{x_q} \right) 
 \psi_n(x_j, \mbf{b}_{\perp j}) \\
  + {\rm (interactions)} . \label{eq:Mb}
 \end{equation}
The normalization is defined by
$\sum_n  \prod_{j=1}^{n-1} \int d x_j d^2 \mathbf{b}_{\perp j}
\left \vert \psi_n(x_j, \mathbf{b}_{\perp j})\right\vert^2 = 1$.
To simplify the discussion we will consider a two-parton hadronic bound state.  In the limit
of zero quark mass
$m_q \to 0$
\begin{equation}  \label{eq:Mbpion}
M^2  =  \int_0^1 \! \frac{d x}{x(1-x)} \int  \! d^2 \mbf{b}_\perp  \,
  \psi^*(x, \mbf{b}_\perp) 
  \left( - \mbf{\nabla}_{ {\mbf{b}}_{\perp}}^2\right)
  \psi(x, \mbf{b}_\perp) +   {\rm (interactions)}.
 \end{equation}

 The functional dependence  for a given Fock state is
given in terms of the invariant mass
\begin{equation}
 M_n^2  = \Big( \sum_{a=1}^n k_a^\mu\Big)^2 = \sum_a \frac{\mbf{k}_{\perp a}^2 +  m_a^2}{x_a}
 \to \frac{\mbf{k}_\perp^2}{x(1-x)} \,,
 \end{equation}
giving  the measure of  the off-energy shell of the bound state,
 $M^2 \! - M_n^2$.
 Similarly in impact space the relevant variable for a two-parton state is  $\zeta^2= x(1-x)\mbf{b}_\perp^2$.
Thus, to first approximation  LF dynamics  depend only on the boost invariant variable
$M_n$ or $\zeta$,
and hadronic properties are encoded in the hadronic mode $\phi(\zeta)$ from the relation
\begin{equation} \label{eq:psiphi}
\psi(x,\zeta, \varphi) = e^{i L \varphi} X(x) \frac{\phi(\zeta)}{\sqrt{2 \pi \zeta}} ,
\end{equation}
thus factoring out the angular dependence $\varphi$ and the longitudinal, $X(x)$, and transverse mode $\phi(\zeta)$
with normalization $ \langle\phi\vert\phi\rangle = \int \! d \zeta \,
 \vert \langle \zeta \vert \phi\rangle\vert^2 = 1$.
 
We can write the Laplacian operator in (\ref{eq:Mbpion}) in circular cylindrical coordinates $(\zeta, \varphi)$
and factor out the angular dependence of the
modes in terms of the $SO(2)$ Casimir representation $L^2$ of orbital angular momentum in the
transverse plane. Using  (\ref{eq:psiphi}) we find~\cite{deTeramond:2008ht}
\begin{equation} \label{eq:KV}  
M^2   =  \int \! d\zeta \, \phi^*(\zeta) \sqrt{\zeta}
\left( -\frac{d^2}{d\zeta^2} -\frac{1}{\zeta} \frac{d}{d\zeta}
+ \frac{L^2}{\zeta^2}\right)
\frac{\phi(\zeta)}{\sqrt{\zeta}}   \\
+ \int \! d\zeta \, \phi^*(\zeta) U(\zeta) \phi(\zeta) ,
\end{equation}
where all the complexity of the interaction terms in the QCD Lagrangian is summed in the effective potential $U(\zeta)$.
The LF eigenvalue equation $P_\mu P^\mu \vert \phi \rangle  =  M^2 \vert \phi \rangle$
is thus a light-front  wave equation for $\phi$
\begin{equation} \label{eq:SLFWE}
\left(-\frac{d^2}{d\zeta^2}
- \frac{1 - 4L^2}{4\zeta^2} + U(\zeta) \right)
\phi(\zeta) = M^2 \phi(\zeta),
\end{equation}
a relativistic single-variable LF Schr\"odinger equation.   Its eigenmodes $\phi(\zeta) = \langle \zeta \vert \phi \rangle$
determine the hadronic mass spectrum and represent the probability
amplitude to find $n$-partons at transverse impact separation $\zeta$,
the invariant separation between pointlike constituents within the hadron~\cite{Brodsky:2006uqa} at equal LF time. 
Extension of the results to arbitrary $n$ follows from the $x$-weighted definition of the
transverse impact variable of the $n-1$ spectator system~\cite{Brodsky:2006uqa}:
$\zeta = \sqrt{\frac{x}{1-x}} \left\vert \sum_{j=1}^{n-1} x_j \mbf{b}_{\perp j} \right\vert$, where $x = x_n$ is the longitudinal 
momentum fraction of the active quark. One can also
generalize the equations to allow for the kinetic energy of massive
quarks using Eqs. (\ref{eq:Mk}) or (\ref{eq:Mb}). In this case, however,
the longitudinal mode $X(x)$ does not decouple from the effective LF bound-state equations.

\section{A Soft-Wall AdS/QCD  Model for  Mesons}

The conformal algebraic structure of  AdS/CFT
can be extended to include a scale $\kappa$. This procedure  breaks conformal
invariance and provides a solution for the confinement of modes, while maintaining an integrable algebraic structure. It also allows one to determine the stability conditions for the solutions. The resulting model resembles the soft wall model 
of Ref.~\cite{Karch:2006pv}.
We write the bound-state LF Hamiltonian as a bilinear product of operators plus a constant $C(\kappa^2)$ to be determined:
\begin{equation} \label{eq:HLC+C}
H_{LF}^\nu(\zeta) = \Pi^\dagger_\nu(\zeta) \Pi_\nu(\zeta)  + C, ~~~ \nu^2 \ge 0,
\end{equation}
where the LF generator  $\Pi$ and its adjoint $\Pi^\dagger$
\begin{equation} \label{eq:PIHO}
\Pi_\nu(\zeta) = -i \left(\frac{d}{d \zeta} - \frac{\nu + \half}{\zeta} 
- \kappa^2 \zeta \right), ~~~~
\Pi^\dagger_\nu(\zeta) = -i \left(\frac{d}{d \zeta} + \frac{\nu + \half}{\zeta}
+\kappa^2 \zeta \right),
\end{equation}
 obey the commutation relation
\begin{equation}
\left[\Pi_\nu(\zeta),\Pi^\dagger_\nu(\zeta)\right] =  \frac{2 \nu+1}{\zeta^2} - 2 \kappa^2.
\end{equation}

For $\nu^2 \ge 0$ and $C \ge - 4 \kappa^2$, the Hamiltonian is positive definite,
$\langle \phi \left\vert H_{LC}^\nu \right\vert \phi \rangle,
= \int d\zeta \, \vert \Pi_\nu \phi(z)  \vert^2+ C \ge 0$ and
$M^2 \ge 0$. For $\nu^2 < 0$ the Hamiltonian cannot be written as
a bilinear product and the Hamiltonian is unbounded from below.
The lowest stable solution of the extended LF Hamiltonian
corresponds to $C = - 4 \kappa^2$ and $\nu = 0$ and it is massless,
$M^2 = 0$.  We impose chiral symmetry by choosing $C = - 4 \kappa^2$ and thus identifying the ground state 
with the pion. With this choice of the constant $C$, the  LF Hamiltonian (\ref{eq:HLC+C}) is
\begin{equation} 
H_{LF}(\zeta) = -\frac{d^2}{d \zeta^2} 
-  \frac{1-4 L^2}{4\zeta^2} + \kappa^4 \zeta^2 + 2 \kappa^2 (L  - 1) ,
\end{equation}
with eigenfunctions
\begin{equation} \label{eq:phiSW}
\phi_L(\zeta) = \kappa^{1+L} \, \sqrt{\frac{2 n!}{(n\!+\!L\!)!}} \, \zeta^{1/2+L}
e^{- \kappa^2 \zeta^2/2} L^L_n(\kappa^2 \zeta^2),
\end{equation}
and eigenvalues $\mathcal{M}^2 = 4 \kappa^2 (n + L)$. This is illustrated in Fig. \ref{mesons} for the pseudoscalar meson spectra.

\begin{figure}
\begin{centering}
\includegraphics[angle=0,width=7.3cm]{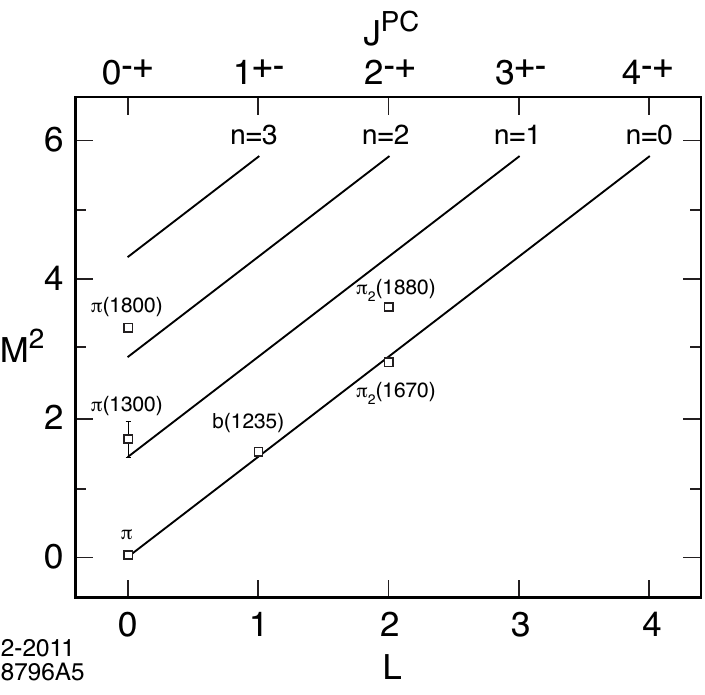} ~~~~
\includegraphics[angle=0,width=7.3cm]{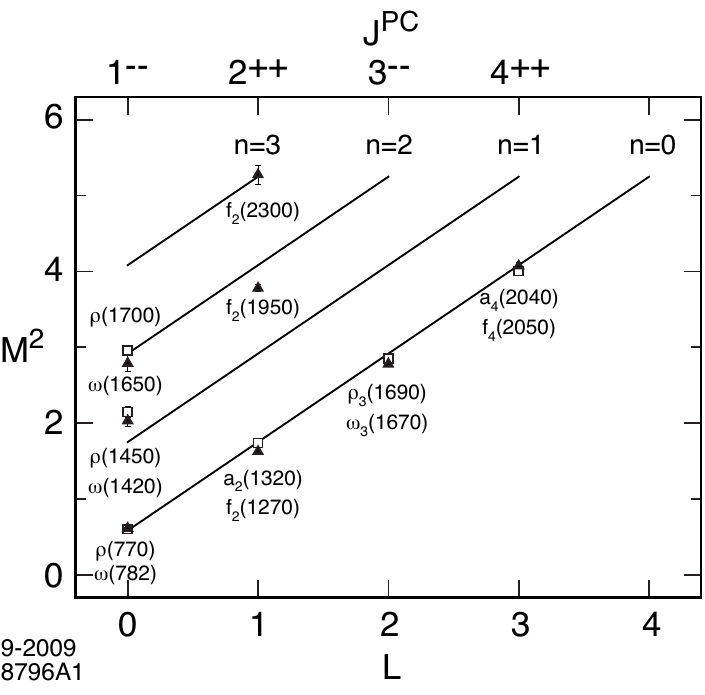}
\caption{\label{mesons} \small Parent and daughter Regge trajectories for the $\pi$-meson family for
$\kappa= 0.6$ GeV (left) and for the  $I\!=\!1$ $\rho$-meson
 and the $I\!=\!0$  $\omega$-meson families for $\kappa= 0.54$ GeV (right). The data are from Ref.~\cite{Amsler:2008xx}.}
\end{centering}
\end{figure}

The confining model has also an effective classical gravity description
corresponding to an AdS$_5$ geometry modified by a positive-sign dilaton background $exp(+ \kappa^2 z^2)$, with sign opposite
to that of reference of Ref.~\cite{Karch:2006pv}.
The positive dilaton solution has interesting physical implications, since  it leads to a confining potential between heavy quarks~\cite{Andreev:2006vy} and to a convenient framework for describing chiral
symmetry breaking.~\cite{Zuo:2009dz}  It also leads to the identification of a nonperturbative effective strong 
coupling $\alpha_s$ and 
$\beta$-functions which are in agreement with available data and  lattice 
simulations.\cite{Brodsky:2010ur}  In the presence of a dilaton profile $exp(+ \kappa^2 z^2)$ the wave equation for a spin $J$ 
mode $\Phi(z) _{\mu_1 \cdots \mu_J}$ is  given by~\cite{deTeramond:2009xk}
\begin{equation} \label{eq:SWJAdS}
\big[ z^2 \partial_z^2 - \left(3 - 2 J - 2 \kappa^2 z^2 \right) z \, \partial_z + z^2 M^2 -  (\mu R)^2 \big] \Phi_J  = 0 .
\end{equation}

Upon the substitution ~$z \! \to\! \zeta$  and  $\phi_J(\zeta)   \!  \sim \! \zeta^{-3/2 + J} e^{\kappa^2 \zeta^2 /2} \, \Phi_J(\zeta)$,
 we find the LF wave equation
\begin{equation}  \label{LFWEJ}
\left(-\frac{d^2}{d \zeta^2} - \frac{1-4 L^2}{4\zeta^2} 
+ \kappa^4 \zeta^2 + 2 \kappa^2(L + S - 1)  \right) \phi_{\mu_1 \cdots \mu_J} 
= M^2 \phi_{\mu_1 \cdots \mu_J},
\end{equation}
with $J_z = L_z + S_z$ and $(\mu R)^2 = - (2-J)^2 + L^2$. Equation  (\ref{LFWEJ}) has eigenfunctions
given by  (\ref{eq:phiSW}) and eigenvalues 
$M_{n, L, S}^2 = 4 \kappa^2 \left(n + L + S/2 \right)$.  The results for $S=1$ vector mesons is illustrated in Fig. \ref{mesons}, where the  spectrum is built by simply adding  $4 \kappa^2$ for a unit change in the radial quantum number, $4 \kappa^2$ for a change in one unit in the orbital quantum number and $2 \kappa^2$ for a change of one unit of spin to the ground state value of $M^2$. Remarkably, the same rule holds for baryons as shown below.

\section{Baryons in Light-Front Holography}

The effective light-front wave equation which describes baryonic states in holographic QCD is a linear equation 
determined by the LF transformation properties of spin 1/2 states.  We write
\begin{equation} \label{eq:LFD}
D_{LF}(\zeta) \psi(\zeta) = M \psi(\zeta),
\end{equation}
where  $D_{LF}$ is a hermitian operator, $D_{LF} = D^\dagger_{LF}$, thus $D_{LF}^2 = M^2$. We write $D_{LF}$ as a
product $D_{LF} = \alpha \Pi$, where $\Pi$ is the matrix valued (non-hermitian) generator
\begin{equation}
\Pi_\nu(\zeta) = -i \left( \frac{d}{d \zeta} - \frac{\nu + \half}{\zeta} \gamma \right) .
\end{equation}
If follows from the square of $D_{LF}$, $D_{LF}^2 = M^2$, that the matrices $\alpha$ and $\gamma$ are $4 \times 4$ anti-commuting hermitian matrices with unit square. The operator
$\Pi$ and its adjoint $\Pi^\dagger$ thus satisfy the commutation relation
\begin{equation}
\left[\Pi_\nu(\zeta),\Pi^\dagger_\nu(\zeta)\right] =  \frac{2 \nu+1}{\zeta^2} \, \gamma.
\end{equation}
The light front Hamiltonian $H_{LF}$ is 
\begin{equation}  
H_{LF}^\nu(\zeta) = \Pi_\nu(\zeta)^\dagger \Pi_\nu(\zeta)  
= - \frac{d^2}{d \zeta^2} 
+ \frac{\left(\nu + \half\right)^2}{\zeta^2} - \frac{\nu + \half}{\zeta^2} \, \gamma.
\end{equation}
The LF equation $H_{LF} \psi_\pm = M^2 \psi_\pm$, has a two-component solution
\begin{equation}
\psi_+(\zeta) \sim \sqrt{\zeta} J_\nu(\zeta M), ~~~~~
\psi_-(\zeta) \sim \sqrt{\zeta} J_{\nu+1}(\zeta M),
\end{equation}
where $\gamma \psi_\pm =  \pm \psi_\pm$. Thus  $\gamma$ is the four dimensional chirality operator $\gamma_5$.
In the Weyl representation
\begin{equation}
\gamma =  
  \begin{pmatrix}
  I&   0\\
  0&  -I
  \end{pmatrix}  ~~~~ {\rm and} ~~~~
 i \alpha =
  \begin{pmatrix}
  0& I\\
- I& 0
  \end{pmatrix}.
  \end{equation}
  
The effective LF equation for baryons (\ref{eq:LFD})   is equivalent to the
Dirac equation describing the propagation of spin-1/2 hadronic modes,  on AdS$_5$ space
$\Psi_P(x^\mu, z) =  e^{-iP \cdot x} \Psi( z)$
\begin{equation} \label{eq:DEz}
\left[i\big( z \eta^{M N} \Gamma_M \partial_N + 2 \, \Gamma_z \big)
 + \mu R \right] \Psi = 0 ,
\end{equation}
where $M, N$ represent the indices of the full space with coordinates  $x^\mu$ and $z$.
Upon  the transformation $\Psi( z) \sim z^2 \psi(z)$, $z \to \zeta$,
 we recover (\ref{eq:LFD})
with $\mu R = \nu + 1/2$ and $\Gamma_z = - i \gamma$.
Higher spin fermionic modes
 $\Psi _{\mu_1 \cdots \mu_{J-1/2}}$, $J > 1/2$, with all polarization indices along the 3+1 coordinates follow by shifting
 dimensions as shown  for the case of mesons.

\section{A Soft-Wall Light-Front Model for Baryons}

An effective LF equation for baryons with a mass gap $\kappa$ is constructed by extending the
conformal algebraic structure for baryons described above, following the analogy with the mesons. 
We write the effective LF Dirac equation
(\ref{eq:LFD}) in terms of the matrix-valued operator $\Pi$ and its adjoint $\Pi^\dagger$ 
\begin{equation} \label{A}
\Pi_\nu(\zeta) = -i\left( \frac{d}{d \zeta} 
- \frac{\nu + \half}{\zeta} \gamma - \kappa^2 \zeta \gamma\right), ~~~
\Pi^\dagger_\nu(\zeta) =  -i\left(\frac{d}{d \zeta} 
+ \frac{\nu + \half}{\zeta} \gamma + \kappa^2 \zeta \gamma\right),
\end{equation}
with the commutation relation
\begin{equation}
\left[\Pi_\nu(\zeta),\Pi^\dagger_\nu(\zeta)\right] =  
\left(\frac{2\nu+1}{\zeta^2} - 2 \kappa^2\right) \gamma.
\end{equation}

The extended  baryonic model also has a geometric interpretation. It corresponds to the Dirac equation in AdS$_5$
space in presence of a linear potential $\kappa^2 z$
\begin{equation} \label{eq:DEz}
\left[i\big( z \eta^{M N} \Gamma_M \partial_N + 2 \, \Gamma_z \big) + \kappa^2 z
 + \mu R \right] \Psi = 0 ,
\end{equation}
as can be shown directly  by using the transformation $\Psi( z) \sim z^2 \psi(z)$, $z \to \zeta$.

As for the case of the mesons
Eq. (\ref{eq:HLC+C})
we write the LF Hamiltonian $H^\nu_{LF} = \Pi_\nu ^\dagger \Pi_\nu + C$ and chose the same value for $C$: $C = - 4 \kappa^2$, effectively modifying the wave equation
(\ref{eq:DEz}).
With this choice for $C$ the LF Hamiltonian is
\begin{equation} \label{eq:LFHs}
H_{LF}= - \frac{d^2}{d \zeta^2} 
+ \frac{\left(\nu + \half\right)^2}{\zeta^2} - \frac{\nu + \half}{\zeta^2} \gamma_5 + \kappa^4\zeta^2 +
\kappa^2 (2 \nu - 3) + \kappa^2 \gamma_5.  
\end{equation}
The LF equation $H_{LF} \psi_\pm = M^2 \psi_\pm$, has a two-component solution
\begin{equation}
\psi_+(\zeta) \sim  \zeta^{\frac{1}{2} + \nu} e^{-\kappa^2 \zeta^2/2}
  L_n^\nu(\kappa^2 \zeta^2) ,\ ~~~
\psi_-(\zeta) \sim  \zeta^{\frac{3}{2} + \nu} e^{-\kappa^2 \zeta^2/2}
 L_n^{\nu+1}(\kappa^2 \zeta^2), 
\end{equation}
and  eigenvalues $M^2 = 4 \kappa^2 (n + \nu)$, identical for both plus and minus eigenfunctions.

\begin{figure}
\begin{centering}
\includegraphics[angle=0,width=14.6cm]{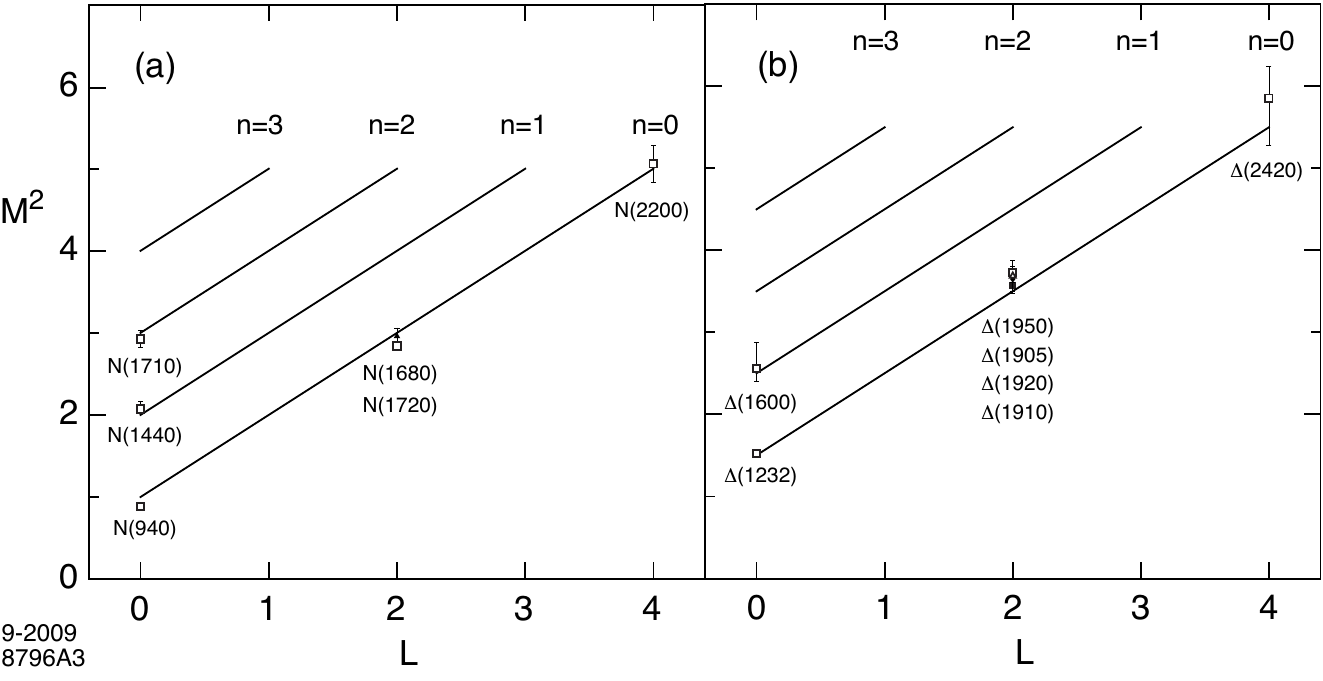}
\caption{\label{baryons}{{\bf 56} \small Parent and daughter Regge trajectories for  the  $N$ and $\Delta$ 
baryon families for $\kappa= 0.5$ GeV.
Data from \cite{Amsler:2008xx}. }
}
\label{baryons}
\end{centering}
\end{figure}

The baryon interpolating operator
$ \mathcal{O}_{3 + L} =  \psi D_{\{\ell_1} \dots
 D_{\ell_q } \psi D_{\ell_{q+1}} \dots
 D_{\ell_m\}} \psi$,  $L = \sum_{i=1}^m \ell_i$, is a twist 3,  dimension $9/2 + L$ operator with scaling behavior given by its
 twist-dimension $3 + L$. We thus require $\nu = L+1$ to match the short distance scaling behavior. Higher spin fermionic modes $\Psi _{\mu_1 \cdots \mu_{J-1/2}}$, $J > 1/2$, are obtained by shifting dimensions for the fields as in the bosonic case. 
Thus, as in the meson sector,  the increase  in the 
mass squared for  higher baryonic states is
$\Delta n = 4 \kappa^2$, $\Delta L = 4 \kappa^2$ and $\Delta S = 2 \kappa^2$, 
relative to the lowest ground state,  the proton.

The predictions for the positive parity light baryons 
 are shown in Fig. \ref{baryons}. As for the predictions for mesons in Fig. \ref{mesons}, only confirmed PDG~\cite{Amsler:2008xx} states are shown. 
The Roper state $N(1440)$ and the $N(1710)$ are well accounted for in this model as the first  and second radial
states. Likewise the $\Delta(1660)$ corresponds to the first radial state of the $\Delta$ family. The model is  successful in explaining the important parity degeneracy observed in the light baryon spectrum, such as the $L\! =\!2$, $N(1680)\!-\!N(1720)$ pair and the $\Delta(1905), \Delta(1910), \Delta(1920), \Delta(1950)$ states which are degenerate 
within error bars. The parity degeneracy of baryons is also a property of the hard wall model, but radial states are not well described by this model.~\cite{deTeramond:2005su} For other
recent calculations of the hadronic spectrum based on AdS/QCD, see Refs.~\cite{Boschi-Filho:2002vd,   BoschiFilho:2005yh, Evans:2006ea, Hong:2006ta, Colangelo:2007pt, Forkel:2007ru, Vega:2008af, Nawa:2008xr, dePaula:2008fp,  Colangelo:2008us, Forkel:2008un, Ahn:2009px, Sui:2009xe, Kirchbach:2010dm, Branz:2010ub}. 

The proton eigenstate in light-front holography
\begin{equation}
\psi(\zeta) = \psi_+(\zeta) u_+  +  \psi_-(\zeta) u_-,
\end{equation}
 has $L^z=0$ and $L^z= + 1$ orbital components combined with spin components $S^z = + 1/2$ and $S^z = -1/2$ respectively. 
An interesting feature of light front holography for baryons and massless quarks is that the lowest valence Fock states with $L^z=0$  and $L^z=\pm 1$ have the same probability
 \begin{equation} \label{intpm}
 \int d \zeta\,  \vert \psi_+(\zeta) \vert^2 = \int d \zeta \, \vert \psi_-(\zeta) \vert^2,
 \end{equation}
 a manifestation of the  chiral invariance of the theory for massless quarks.~\cite{Brodsky:2010px}
 This implies that the quarks carry zero angular momentum $\langle S_z=0\rangle$ in the proton with $J^z=\pm 1/2$ and $\langle L^z=1/2\rangle$.

There are many other interesting predictions for baryons using AdS/QCD and light front holographic methods such as space-like and time-like Dirac and Pauli form factors, valence structure functions, etc.  The AdS/QCD LF Hamiltonian also creates Fock states with extra quark-antiquark pairs~\cite{deTeramond:2010ez} as in QCD(1+1).

\section{Conclusions}

We have derived a correspondence between a semiclassical first approximation to QCD quantized on the light-front
and hadronic modes propagating on a fixed AdS background. This provides a duality between the bosonic
and fermionic wave equations in AdS higher dimensional space  and the corresponding LF equations in
physical 3 + 1 space.  
The duality leads to  Schr\"odinger and Dirac-like equations  for 
hadronic bound states  in physical space-time when one identifies the 
AdS fifth dimension coordinate $z$ with the LF coordinate $\zeta$.
The light-front equations of motion, which are dual to an effective classical gravity theory, possess remarkable algebraic and integrability properties which follow from the underlying conformal properties of the theory.  We also extend the algebraic construction  to include a confining potential while preserving  the integrability of the mesonic and baryonic bound-state equations.

Light-Front Holography is one of the most remarkable features of AdS/CFT.  It  allows one to project the functional dependence of the wavefunction $\Phi(z)$ computed  in the  AdS fifth dimension to the  hadronic frame-independent light-front wavefunction $\psi(x_i, \mbf{b}_{\perp i})$ in $3+1$ physical space-time. The 
variable $z $ maps  to $ \zeta(x_i, \mbf{b}_{\perp i})$. To confirm this, we have shown that there exists a correspondence between the matrix elements of the energy-momentum tensor of the fundamental hadronic constituents in QCD with the transition amplitudes describing the interaction of string modes in anti-de Sitter space with an external graviton field which propagates in the AdS interior. The agreement of the results for both electromagnetic and gravitational hadronic transition amplitudes provides an important consistency test and verification of holographic mapping from AdS to physical observables defined on the light-front. As we have discussed, this correspondence is a consequence of the fact that the metric $ds^2$ for AdS$_5$ at fixed light-front time $\tau$ is invariant under the simultaneous scale change  $\mbf{x}^2_\perp \to \lambda^2 \mbf{x}^2_\perp $ in transverse space and $z^2 \to \lambda^2 z^2$.  The transverse coordinate $\zeta$ is closely related to the invariant mass squared  of the constituents in the LFWF  and its off-shellness  in  the light-front kinetic energy,  and it is thus the natural variable to characterize the hadronic wavefunction.  In fact $\zeta$ is the only variable to appear in the light-front
Schr\"odinger equations predicted from AdS/QCD.  These equations for both meson and baryons give a good representation of the observed hadronic spectrum. The resulting LFWFs also have remarkable phenomenological features, including predictions for the  electromagnetic form factors and decay constants.  We have also shown that the LF Hamiltonian formulation of quantum field theory provides a natural formalism to compute 
hadronization at the amplitude level.~\cite{Brodsky:2008tk}

The light-front  holographic theory provides successful predictions for the light-quark meson and baryon spectra, as function of hadron spin, quark angular momentum, and radial quantum number. Using the positive dilaton background $\exp( + \kappa^2 z^2)$  the pion is massless, corresponding to zero mass quarks, in agreement with chiral invariance arguments. Higher spin light-front equations can be derived by shifting dimensions in the AdS wave equations.~\cite{deTeramond:2009xk}
Unlike the top-down string theory approach,  one is not limited to hadrons of maximum spin
$J \le 2$, and one can study baryons with finite color $N_C=3.$
Both the hard and soft-wall models  predict similar multiplicity of states for mesons
and baryons as it is observed experimentally.~\cite{Klempt:2007cp}
In the hard-wall model the dependence  has the form:  $M \sim 2n + L$. 
However, in the soft-wall
model the observed  Regge behavior is found: $M^2 \sim n + L,$  which has the same slope in radial quantum number and orbital angular momentum.

The semiclassical AdS/QCD approximation to light-front QCD 
described in this talk breaks down at short distances
where hard gluon exchange and quantum corrections become important. 
However, one can systematically
improve the semiclassical approximation by introducing nonzero quark masses and short-range Coulomb
corrections,  thus extending the predictions of the model to the dynamics and spectra of heavy and heavy-light quark systems.  One can also diagonalize the LF Hamiltonian as in DLCQ, but  on the orthonormal basis states provided by AdS/QCD.~\cite{Vary:2009gt} One could also employ Lippmann-Schwinger perturbation theory, systematically correcting the AdS/QCD eigensolutions.

\section*{Acknowledgments}

Presented by SJB at the International Conference on the Structure of Baryons\\
BARYONS'10, December  7-11, 2010, Osaka, Japan.  This research was supported by the Department
of Energy contract DE--AC02--76SF00515. SLAC-PUB-14381


\begin{thebibliography}{99}



\bibitem{Dirac:1949cp}
  P.~A.~M.~Dirac,
  Rev.\ Mod.\ Phys.\  {\bf 21}, 392 (1949).
  
  
\bibitem{Brodsky:1997de}
  For a review of light-front quantization, 
   see S.~J.~Brodsky, H.~-C.~Pauli, S.~S.~Pinsky,
  Phys.\ Rept.\  {\bf 301}, 299-486 (1998)
  [hep-ph/9705477].
  
\bibitem{Brodsky:2008tk}
  S.~J.~Brodsky, G.~de Teramond, R.~Shrock,
  AIP Conf.\ Proc.\  {\bf 1056}, 3-14 (2008)
  [arXiv:0807.2484 [hep-ph]].
  
\bibitem{Munger:1993kq}
  C.~T.~Munger, S.~J.~Brodsky, I.~Schmidt,
  Phys.\ Rev.\  {\bf D49}, 3228-3235 (1994).


\bibitem{Brodsky:1980pb}
  S.~J.~Brodsky, P.~Hoyer, C.~Peterson, N.~Sakai,
  Phys.\ Lett.\  {\bf B93}, 451-455 (1980).
  
\bibitem{Bari:1991ty}
  G.~Bari, M.~Basile, G.~Bruni {\it et al.},
  Nuovo Cim.\  {\bf A104}, 1787-1800 (1991).


\bibitem{Mattson:2002vu}
  M.~Mattson {\it et al.} [ SELEX Collaboration ],
  Phys.\ Rev.\ Lett.\  {\bf 89}, 112001 (2002)
  [hep-ex/0208014].
  
  \bibitem{Lansberg}
  S.~J.~Brodsky, .J~Fleuret,  J.~P.~Lansberg, C.~Lorce (in progress).
 
  
\bibitem{Brodsky:2006wb}
  S.~J.~Brodsky, B.~Kopeliovich, I.~Schmidt,  J.-Soffer, 
  Phys.\ Rev.\  {\bf D73}, 113005 (2006)
  [hep-ph/0603238].
  
\bibitem{Brodsky:2007yz}
  S.~J.~Brodsky, A.~S.~Goldhaber, B.~Z.~Kopeliovich,I.~Schmidt,
  Nucl.\ Phys.\  {\bf B807}, 334-347 (2009)
  [arXiv:0707.4658 [hep-ph]].

\bibitem{Arleo:2009ch}
  F.~Arleo, S.~J.~Brodsky, D.~S.~Hwang, A.~Sickles,
  Phys.\ Rev.\ Lett.\  {\bf 105}, 062002 (2010)
  [arXiv:0911.4604 [hep-ph]].
  
\bibitem{Brodsky:2008qp}
  S.~J.~Brodsky, A.~Sickles,
  Phys.\ Lett.\  {\bf B668}, 111-115 (2008)
  [arXiv:0804.4608 [hep-ph]].



\bibitem{Brodsky:2002cx}
  S.~J.~Brodsky, D.~S.~Hwang, I.~Schmidt,
  Phys.\ Lett.\  {\bf B530}, 99-107 (2002)
  [hep-ph/0201296].
  
\bibitem{Collins:2002kn}
  J.~C.~Collins,
  Phys.\ Lett.\  {\bf B536}, 43-48 (2002)
  [hep-ph/0204004].
  
\bibitem{Brodsky:2002rv}
  S.~J.~Brodsky, D.~S.~Hwang, I.~Schmidt,
  Nucl.\ Phys.\  {\bf B642}, 344-356 (2002)
  [hep-ph/0206259].

\bibitem{Boer:2002ju}
  D.~Boer, S.~J.~Brodsky, D.~S.~Hwang,
  Phys.\ Rev.\  {\bf D67}, 054003 (2003).
  [hep-ph/0211110].
  
  
\bibitem{Brodsky:2009zd}
  S.~J.~Brodsky, R.~Shrock,
  Proc.\ Nat.\ Acad.\ Sci.\  {\bf 108}, 45-50 (2011)
  [arXiv:0905.1151 [hep-th]].

\bibitem{Brodsky:2010xf}
  S.~J.~Brodsky, C.~D.~Roberts, R.~Shrock , P.~Tandy,
  Phys.\ Rev.\  {\bf C82}, 022201 (2010)
  [arXiv:1005.4610 [nucl-th]].


\bibitem{Maldacena:1997re}
  J.~M.~Maldacena,
  Adv.\ Theor.\ Math.\ Phys.\  {\bf 2}, 231 (1998)
  [Int.\ J.\ Theor.\ Phys.\  {\bf 38}, 1113 (1999)]
  [arXiv:hep-th/9711200].

\bibitem{deTeramond:2008ht}
  G.~F.~de Teramond and S.~J.~Brodsky,
  Phys.\ Rev.\ Lett.\  {\bf 102}, 081601 (2009)
  [arXiv:0809.4899 [hep-ph]].

\bibitem{Brodsky:2006uqa}
  S.~J.~Brodsky and G.~F.~de Teramond,
  Phys.\ Rev.\ Lett.\  {\bf 96}, 201601 (2006)
  [arXiv:hep-ph/0602252];
  Phys.\ Rev.\  {\bf D77}, 056007 (2008)
  [arXiv:0707.3859 [hep-ph]].

\bibitem{Brodsky:2008pf}
  S.~J.~Brodsky and G.~F.~de Teramond,
  Phys.\ Rev.\  D {\bf 78}, 025032 (2008)
  [arXiv:0804.0452 [hep-ph]].

\bibitem{Karch:2006pv}
  A.~Karch, E.~Katz, D.~T.~Son and M.~A.~Stephanov,
  Phys.\ Rev.\  D {\bf 74}, 015005 (2006)
  [arXiv:hep-ph/0602229].


  \bibitem{Amsler:2008xx}
  C. Amsler {\it et al.}  (Particle Data Group),
  Phys.\ Lett.\  B {\bf 667} (2008) 1.


\bibitem{Andreev:2006vy}
  O.~Andreev,
  Phys.\ Rev.\  D {\bf 73}, 107901 (2006)
  [arXiv:hep-th/0603170].

\bibitem{Zuo:2009dz}
  F.~Zuo,
  Phys.\ Rev.\  D {\bf 82}, 086011 (2010)
  [arXiv:0909.4240 [hep-ph]].

\bibitem{Brodsky:2010ur}
  S.~J.~Brodsky, G.~F.~de Teramond, A.~Deur,
  Phys.\ Rev.\  {\bf D81}, 096010 (2010)
  [arXiv:1002.3948 [hep-ph]].
  
  \bibitem{deTeramond:2009xk}
  G.~F.~de Teramond, S.~J.~Brodsky,
  Nucl.\ Phys.\ Proc.\ Suppl.\  {\bf 199}, 89-96 (2010)
  [arXiv:0909.3900 [hep-ph]].

  
  \bibitem{deTeramond:2005su}
  G.~F.~de Teramond and S.~J.~Brodsky,
  Phys.\ Rev.\ Lett.\  {\bf 94}, 201601 (2005)
  [arXiv:hep-th/0501022].
  
   \bibitem{Boschi-Filho:2002vd}
  H.~Boschi-Filho and N.~R.~F.~Braga,
  JHEP {\bf 0305}, 009 (2003)
  [arXiv:hep-th/0212207].


\bibitem{BoschiFilho:2005yh}
  H.~Boschi-Filho, N.~R.~F.~Braga and H.~L.~Carrion,
  Phys.\ Rev.\  D {\bf 73}, 047901 (2006)
  [arXiv:hep-th/0507063].

\bibitem{Evans:2006ea}
  N.~Evans and A.~Tedder,
  Phys.\ Lett.\  B {\bf 642}, 546 (2006)
  [arXiv:hep-ph/0609112].

\bibitem{Hong:2006ta}
  D.~K.~Hong, T.~Inami and H.~U.~Yee,
  Phys.\ Lett.\  B {\bf 646}, 165 (2007)
  [arXiv:hep-ph/0609270].

\bibitem{Colangelo:2007pt}
  P.~Colangelo, F.~De Fazio, F.~Jugeau and S.~Nicotri,
  Phys.\ Lett.\  B {\bf 652}, 73 (2007)
  [arXiv:hep-ph/0703316].

\bibitem{Forkel:2007ru}
  H.~Forkel,
  Phys.\ Rev.\  D {\bf 78}, 025001 (2008)
  [arXiv:0711.1179 [hep-ph]].

\bibitem{Vega:2008af}
  A.~Vega and I.~Schmidt,
  Phys.\ Rev.\  D {\bf 78}, 017703 (2008)
  [arXiv:0806.2267 [hep-ph]].

\bibitem{Nawa:2008xr}
  K.~Nawa, H.~Suganuma and T.~Kojo,
  Mod.\ Phys.\ Lett.\  A {\bf 23}, 2364 (2008)
  [arXiv:0806.3040 [hep-th]].

\bibitem{dePaula:2008fp}
  W.~de Paula, T.~Frederico, H.~Forkel and M.~Beyer,
  Phys.\ Rev.\  D {\bf 79}, 075019 (2009)
  [arXiv:0806.3830 [hep-ph]].

\bibitem{Colangelo:2008us}
  P.~Colangelo, F.~De Fazio, F.~Giannuzzi, F.~Jugeau and S.~Nicotri,
  Phys.\ Rev.\  D {\bf 78}, 055009 (2008)
  [arXiv:0807.1054 [hep-ph]].

\bibitem{Forkel:2008un}
  H.~Forkel and E.~Klempt,
  Phys.\ Lett.\  B {\bf 679}, 77 (2009)
  [arXiv:0810.2959 [hep-ph]].

\bibitem{Ahn:2009px}
  H.~C.~Ahn, D.~K.~Hong, C.~Park and S.~Siwach,
  Phys.\ Rev.\  D {\bf 80}, 054001 (2009)
  [arXiv:0904.3731 [hep-ph]].

\bibitem{Sui:2009xe}
  Y.~Q.~Sui, Y.~L.~Wu, Z.~F.~Xie and Y.~B.~Yang,
  Phys.\ Rev.\  D {\bf 81}, 014024 (2010)
  [arXiv:0909.3887 [hep-ph]].
  
  \bibitem{Kirchbach:2010dm}
  M.~Kirchbach, C.~B.~Compean,
  Phys.\ Rev.\  {\bf D82}, 034008 (2010)
  [arXiv:1003.1747 [hep-ph]].
  
  
\bibitem{Branz:2010ub}
  T.~Branz, T.~Gutsche, V.~E.~Lyubovitskij {\it et al.},
  Phys.\ Rev.\  {\bf D82}, 074022 (2010)
  [arXiv:1008.0268 [hep-ph]].
 
 \bibitem{Brodsky:2010px}
  S.~J.~Brodsky, G.~F.~de Teramond,
  Acta Phys. Pol. B 41, 2605 (2010)
  [arXiv:1009.4232 [hep-ph]].

\bibitem{deTeramond:2010ez}
  G.~F.~de Teramond, S.~J.~Brodsky
  [arXiv:1010.1204 [hep-ph]].
  

  \bibitem{Klempt:2007cp}
  E.~Klempt and A.~Zaitsev,
  Phys.\ Rept.\  {\bf 454}, 1 (2007)
  [arXiv:0708.4016 [hep-ph]].
     
   \bibitem{Vary:2009gt}
  J.~P.~Vary, H.~Honkanen, J.~Li {\it et al.},
  Phys.\ Rev.\  {\bf C81}, 035205 (2010)
  [arXiv:0905.1411 [nucl-th]].

  
   
\end{thebibliography}
\end{document}